\documentstyle{article}
\begin{document}

\title{The first BritGrav meeting, \break
Southampton, 27/28 March 2001}

\maketitle

\parindent 0 mm
\parskip 1 mm

Many relativists now working in Britain have good memories of the
Pacific Coast, Midwest or Nickel and Dime meetings from their postdocs
in the US. So it seemed natural to establish a similar annual meeting
in Britain, and the Southampton GR group gave it a try.

The two distinguishing features of the US regional meetings are very
short talks, and keeping it simple and cheap. This concept proved a
success east of the Atlantic too: 81 people attended, giving 47
10-minute plenary talks over the two days. 12 of the talks were by
PhD students, and 8 by postdocs: a proportion we hope to increase in
the future!

On the two days, talks were only roughly grouped by subject. The
distribution of topics differed noticeably from those of recent US
regional meetings. Below are the abstracts of all talks in the order
in which they were given. (The electronic preprint references have
been added by the organizers at the request of xxx admin, and are
indicative only.)

The {\bf BritGrav02} meeting will be organised by Henk van Elst and
Reza Tavakol at Queen Mary and Westfield College, London. All
enquiries to them: \hfill\break 
H.van.Elst@qmw.ac.uk and r.tavakol@maths.qmw.ac.uk.

\begin{enumerate}

\item Carlos Sopuerta, U Portsmouth, carlos.sopuerta@port.ac.uk

Dynamics of irrotational dust matter in the long wavelength approximation
       
We report results on the long wavelength iteration of the general
relativistic equations for irrotational dust matter in the covariant
fluid approach.  In particular, we discuss the dynamics of these
models during the approach to any spacelike singularity where a
BKL-type evolution is expected, studying the validity of this
approximation scheme and the role of the magnetic part of the Weyl
tensor.

\item  Spyros S Kouris, U York, ssk101@york.ac.uk

Large-distance behavior of graviton two-point functions in de Sitter
spacetime (gr-qc/0004097)

It has been observed that the graviton two-point functions in de
Sitter spacetime in various gauges grow as the distance between the
two points increases.  We show that this behavior is a gauge artifact
in a non-covariant gauge.  We argue that it is also a gauge artifact
in a two-parameter family of covariant gauges.  In particular, we show
that the two-point function of the linearized Weyl tensor is
well-behaved at large distances.

\item Stanislav Babak, U Cardiff, Stanislav.Babak@astro.cf.ac.uk

Finite-range gravity and its role in cosmology, black holes and
gravitational waves

The Field Theoretical approach to gravity provides us with a natural way
to modify general relativity. In this paper we have considered a two parameter
family of theories of a finite-range gravitational field. To give a proper
physical interpretation, we have considered the exact solutions of
linearised equations. They describe plane gravitational waves and static
spherically symmetric gravitational field. A certain choice of sign of the
free parameters allows us to associate these free parameters with the rest
masses of longitudinal and transverse gravitons. In the static and spherically
symmetric problem we have obtained  Yukawa-type gravitational potentials
instead of Coulomb-type and the gravitational field becomes finite ranged.
Applying the theory of finite-range gravitational field to the homogeneous and
isotropic Universe, we have shown that even a very small mass of longitudinal
graviton can drastically alter the late-time evolution of the Universe.
According to the sign of the free parameters, the expansion of the Universe
either slows down or gains an additional acceleration. Numerical and
semi-analytical solutions of the exact field equations for the static and
spherically symmetric problem (Schwarzschild-like solution) were obtained.  
It has been demonstrated that the event horizon occurs at the location of
the physical singularity. That is, a regular event horizon is unstable with
respect to ascribing graviton with a non-zero rest mass.

\item Cristiano Germani, U Portsmouth, Cristiano.Germani@port.ac.uk

Gravitational collapse in the brane

Abstract: We discuss some aspects of gravitational collapse in the brane
world scenario, focusing on the 4-D brane and new features arising from the
modified Einstein equations. In particular we report results on collapse of
a pure Weyl field and the possible formation of a pure Weyl-charged  black
hole whose metric is formally that of  Reissner-Nordstrom, but  with no mass
and a negative charge term.

\item Jorma Louko, U Nottingham, jorma.louko@maths.nottingham.ac.uk

Brane worlds with bolts

We construct a brane-world model that has one compact extra dimension
on the brane, two extra dimensions in the bulk, and a nonvanishing
bulk magnetic field. The main new feature is that the bulk has no
horizons that could develop singularities upon the addition of
perturbations. The static scalar propagator is calculated on the brane
and shown not to see the extra dimensions in the large distance limit.
We argue, in part on grounds of an exact nonlinear gravitational wave
solution on the brane-world background, that a similar result should
hold for linearised gravity.

\item  M.L. Fil'chenkov, Peoples' Friendship University, Moscow, fil@agmar.ru

Tunnelling models of creation and collapse

The early Universe and late collapse are considered in terms of quantum 
tunnelling through some potential barrier constructed from Einstein's 
equations. Wave functions, energy levels and a penetration factor are 
calculated for these quantum systems. Applications to creation of a 
universe in the laboratory, observational cosmology and miniholes are 
discussed. A possibility of the creation of open and flat models  as 
well as a role of quintessence (de Sitter vacuum, domain walls and 
strings) in these processes are investigated. 

\item Henk van Elst, Queen Mary and Westfield College,
henk@gmunu.mth.uct.ac.za

Scale-invariant dynamics for Abelian G2 perfect fluid cosmologies

A dynamical formulation at a derivative level $\partial^{2}g$ for
Abelian G2 perfect fluid cosmologies is introduced that employs
scale-invariant autonomous evolution systems of symmetric hyperbolic
format. This allows for a transparent isolation of (i) the physical
degrees of freedom in both the gravitational and the matter source
fields and (ii) the gauge degrees of freedom associated with the time
slicing. In addition, the self-similar (asymptotic) states can be
determined systematically. Various applications are highlighted.

\item Sonny Khan, U Aberdeen, S.Khan@maths.abdn.ac.uk

Projective symmetries in space-times

The existence of proper projective vector fields is discussed in
Einstein-Maxwell and spherically symmetric static space-times.  The
problem is resolved for null Einstein-Maxwell space-times (where none
can exist) and under certain restrictions in the non-null case (where
none have so far been found to exist).  Examples of such vector fields
are provided in spherically symmetric static space-times, where a
general solution (under a loose restriction) is presented.

\item Ghulam Shabbir, U Aberdeen, shabbir@maths.abdn.ac.uk

Curvature collineations for certain space-time metrics

A approach is suggested using the 6x6 form of the curvature tensor to
find the complete set of curvature collineations (CCs) in space- times
which possess certain types of (metric) symmetry. This approach
immediately rules out the possibilities where proper CCs cannot exist
and suggests how to find CCs when they do. The space-times considered
include those with plane and spherical symmetric, static symmetry.

\item Graham Hall, U Aberdeen, gsh@maths.abdn.ac.uk

Orbits of symmetries in space-times

The subject of the orbits of symmetries in general relativity is
usually discussed in an ad-hoc way. The object of this talk is to try 
to clarify the position and to offer precise definitions and outline
rigorous proofs. The questions to be answered (totally or partially) 
include

(i) exactly what are these symmetries, how are they and their orbits
    described and in what sense,if any, are they groups?

(ii) how do the standard geometrical invariants behave on an orbit?

(iii)what are "fixed points" of symmetries, what happens there and where
     can they occur?  

(iv) are there "well behaved" and "badly behaved" orbits and if so, 
     how does one distinguish between them and do the well behaved ones
   behave as in the "folklore" of the subject?

(v)  what rules determine the dimension and type of an orbit once the
     symmetries are specified?

\item Raul Vera, Queen Mary and Westfield College, R.Vera@qmw.ac.uk

Matching preserving the symmetry

In the literature, the matchings between spacetimes have been most of
the times implicitly assumed to preserve the symmetry. But no
definition for such a kind of matching was given until very
recently. Loosely speaking, the matching hypersurface is restricted to
be tangent to the orbits of a desired group of isometries admitted at
both sides of the matching and thus admitted by the whole matched
spacetime. This restriction can lead to conditions on the properties
of the preserved group of isometries, such as its algebraic type and
the geometrical xproperties of the vector fields that generate that
group.

\item Bill Bonnor, QMW, 100571.2247@compuserve.com

Equilibrium of classical spinning particles

Using an approximation method I investigate the stationary axisymmetric
solution for two spinning mass particles.  It contains, as expected, a
conical singularity between the particles representing a strut preventing
collapse.  However, there is a second singularity which seems to represent a
torque preserving the spins of the particles.  For certain
values of the spins no torque is needed.

It does not seem possible to explain this solution in terms of classical
mechanics.

\item Alan Barnes, Aston University, barnes@aston.ac.uk

On some perfect fluid solutions of Stephani

Some years ago Stephani derived several solutions for a geodesic perfect
fluid flow with constant pressure.  In this paper Stephani's solutions
with non-zero rotation are generalised;  all are of Petrov type D and
the magnetic part of the Weyl tensor vanishes. In general the solutions
admit no Killing vectors and the fluid flow is shearing, twisting and
expanding. The solutions can all be matched across a time-like
hypersurface of constant curvature to a de Sitter or Minkowski spacetime.

A generalisation of Stephani's ansatz is also considered.  The general
solution in this case has not yet been derived, but some very simple
exact solutions for a fluid with spherical symmetry have been obtained.

\item Brian Edgar, U Linkoping, bredg@mai.liu.se

Tetrads and symmetry

Two of the main successful tools in the search for exact solutions of
Einstein's equations are tetrad formalisms and symmetry
groups. However, when these two methods are used together there is a
lot of redundancy in the calculations, and the two methods do not
complement each other. Chinea, Collinson and Held have in the past
looked at the possibility of introducing symmetry conditions at tetrad
level, and more recently Fayos and Sopuerto have proposed a new
approach to integrating tetrads and symmetry.  Building on the results
of Chinea, Collinson and Held we propose and illustrate a method which
analyses easily and efficiently the Killing vector structure of
metrics as they are calculated in tetrad formalisms.

\item Fredrik Andersson, U Linkoping frand@mai.liu.se

Potentials and superpotentials of symmetric spinor fields

In 1988 Illge proved that an arbitrary symmetric (n,0)-spinor field
always has an (n-1,1)-spinor potential, which is symmetric over its n-1
unprimed indices. In particular this gives an alternative proof for the
existence of a Lanczos potential of the Weyl spinor. Illge also
considered the problem of finding completely symmetric spinor potentials
for completely symmetric spinor fields having both primed and unprimed
indices. Because of algebraic inconsistencies it turned out to be
impossible to prove a general existence theorem for these potentials.
However, in one important special case it is possible to prove existence
of completely symmetric spinor potentials. In an Einstein spacetime it
turns out that if we look for potentials for spinors having only one
primed index, the algebraic inconsistencies collapse into differential
conditions which can be satisfied using gauge freedom. Thus, in Einstein
spacetimes, completely symmetric (n,1)-spinor fields always has a
completely symmetric (n-1,2)-spinor potential. This means that the Weyl
spinor of an Einstein spacetime always has a completely symmetric (2,2)-
spinor potential. This 'superpotential' seems to be related to quasi-local
momentum of the Einstein spacetime.

\item Annelies Gerber, Imperial College, annelies.gerber@ic.ac.uk, and
Patrick Dolan, Imperial College, pdolan@inctech.com

The Lanczos curvature potential problems with applications

The Weyl- and Riemann curvature tensors have both been analysed in
terms of a tensor potential $L_{abc}$. The Weyl-Lanczos system of PDE's
is always in involution but the Riemann-Lanczos system needs
prolongation to be in involution in general.  Examples to illustrate
these problems are given.

\item Robin Tucker, U Lancaster, r.tucker@lancaster.ac.uk

On the detection of scalar field induced spacetime torsion (gr-qc/0104050)

It is argued that the geodesic hypothesis based on autoparallels of
the Levi-Civita connection may need refinement in the Brans-Dicke
theory of gravitation. Based on a reformulation of this theory in
terms of a connection with torsion determined dynamically in terms of
the gradient of the Brans-Dicke scalar field, we compute the
perihelion shift in the orbit of Mercury on the alternative hypothesis
that its worldline is an autoparallel of a connection with torsion. If
the Brans-Dicke scalar field couples significantly to matter and test
particles move on such worldlines, the current time keeping methods
based on the conventional geodesic hypothesis may need refinement.

\item Julian Barbour, jbarbour@online.rednet.co.uk

Relativity without relativity (gr-qc/0012089)

I shall give a brief review of the above paper by myself and Brendan
Foster and Niall Ó Murchadha. We give a new derivation of general
relativity based entirely on three dimensional principles. We start
with a parametrisation invariant, Jacobi-type action on
superspace. This will be the product of a square root of a potential
times the square root of a kinetic energy term. All we demand is that
the action have nontrivial solutions. We find that the only viable
action is the Baierlein-Sharp-Wheeler Lagrangian and thus we recover
G.R. We impose no spacetime conditions whatsoever. We extend this to
include scalar and vector fields. We recover causality (everything
travels at the same speed), Maxwellian electrodynamics, and the gauge
principle. Thus we derive a large part of modern physics from a purely
three dimensional point of view. (gr-qc/0012089)

\item Petros Florides, Trinity College Dublin, florides@maths.tcd.ie

The  Sagnac  effect  and  the  special theory  of  relativity

Contrary to the recent claim by Dr A.G. Kelly and Professor
J.P. Vigier, it is shown, in two distinct ways, that the Sagnac Effect
and Special Relativity are in complete and perfect harmony.

\item John Barrett, U Nottingham, John.Barrett@nottingham.ac.uk

Quantum Gravity and the Lorentz group

I will give a brief summary of the state of progress of models of 4d
quantum gravity based on the representation theory of the Lorentz group,
and its future prospects.

\item  Christopher Steele, U Nottingham,
   Christopher.Steele@maths.nottingham.ac.uk

Asymptotics of relativistic spin networks

I will discuss Relativistic Spin Networks based on the representation
theory of the 4 dimensional rotation group. I will present asymptotic
formulae for the evaluation of particular networks and provide a
geometrical interpretation.

\item Robert Low, U Coventry, mtx014@coventry.ac.uk

Timelike foliations and the shape of space

What is the shape of space in a space-time? In the familar
case of a globally hyperbolic space-time, one natural answer
is to consider the topology of a Cauchy surface. However,
there are other approaches which one might also consider.
One is to consider edgeless spacelike submanifolds of
the space-time; another is to foliate the space-time
by timelike curves, and consider the quotient space 
obtained by identifying points lying on the same curve.
I will describe conditions on the family of timelike curves,
and on a vector field whose integral curves they are
for this to give rise to a meaningful shape of space,
and briefly discuss the relationship between this approach
and that of considering edgeless spacelike submanifolds.

\item Jonathan Wilson, U Southampton, jpw@maths.soton.ac.uk

Generalised hyperbolicity in singular space-times (gr-qc/0001079,
gr-qc/0101018) 

A desirable property of any physically plausible space-time is global
hyperbolicity. It is shown that a weaker form of hyperbolicity, defined
according to whether the scalar wave equation admits a unique solution, 
is satisfied in certain space-times with weak singularities such as those
containing thin cosmic strings or shells of matter.  It therefore evident
that such weak singularities may be regarded as internal points of
space-time.

\item Rod Halburd, U Loughborough, R.G.Halburd@lboro.ac.uk

Painleve analysis in General Relativity
Speaker:  Rod Halburd, Dept Mathematical Sciences, Loughborough University

Painleve analysis uses the singularity structure of solutions of a
differential equation in the complex domain as an indicator of the
integrability (solvability) of a differential equation.  A large class
of charged spherically symmetric models will be identified and solved
using this method.

\item Magnus Herberthson, U Linköping, maher@mai.liu.se

A nice differentiable structure at spacelike infinity (gr-qc/9712058)

By a conformal rescaling and compactification of the (asymptotically
flat) physical space-time, spacelike infinity is represented by a
single point. It it known that the regularity of the manifold at that
point cannot be smooth, and various differentiable structures have
been suggested. In this talk we report that in the case of a Kerr
solution, the standard C>1-structure can be extended to include both
spacelike and null directions from spacelike infinity.

\item Jonathan Thornburg, U Vienna, jthorn@thp.univie.ac.at

Episodic self-similarity in critical gravitational collapse (gr-qc/0012043)

I report on a new behavior found in numerical simulations of spherically
symmetric gravitational collapse in self-gravitating \hbox{SU(2)}
$\sigma$~models at intermediate gravitational coupling constants:
The critical solution (between black hole formation and dispersion)
closely approximates the continuously self-similar (CSS) solution for
a finite time interval, then departs from this, and then returns to
CSS again.  This cycle repeats several times, each with a different
CSS accumulation point.  The critical solution is also approximately
discretely self-similar (DSS) throughout this whole process.

\item Jose Maria Martin Garcia, U Southampton, jmm@maths.soton.ac.uk

Stability of Choptuik spacetime in the presence of charge and
angular momentum.

We show that Choptuik spacetime is a codimension-1 exact solution of
the full Einstein - Maxwell - Klein-Gordon problem. That is,
electromagnetic field perturbations, charged scalar perturbations and
perturbations with angular momentum all decay. Only the well known
spherical neutral perturbation linking Chopuik spacetime with
Schwarzschild and Minkowski is unstable. We calculate critical
exponents for charge and angular momentum for near critical collapse.

\item Elizabeth Winstanley, U Sheffield, e.winstanley@sheffield.ac.uk

Update on stable hairy black holes in AdS

Black holes in anti-de Sitter space can support gauge field hair 
which is stable under spherically symmetric perturbations.  
This talk discusses recent work showing that these black holes 
remain stable under non-spherically symmetric perturbations in the 
odd-parity sector.

\item VS Manko, CINVESTAV - IPN, VladimirS.Manko@fis.cinvestav.mx

Equilibrium configurations of aligned black holes

The existence of multi-black hole equilibrium configurations in
different axisymmetric systems is discussed.

\item  Colin Pendred, U Nottingham, Colin.Pendred@maths.nottingham.ac.uk

Black hole formation in (2+1)-dimensional relativity

The non-spinning BTZ black hole is introduced and shown that it can be
formed by the collision of two point particles in (2+1)-dimensional
spacetimes of negative cosmological constant. The more general,
spinning BTZ black hole is then considered.

\item Atsushi Higuchi, U York, ah28@york.ac.uk

Low-energy absorption cross sections of stationary black holes (gr-qc/0011070)

We present a special-function free derivation of the fact shown first
by Das, Gibbons and Mathur that the low-energy massless scalar
absorption cross section of a spherically symmetric black hole is
universally given by the horizon area. Our derivation seems to
generalize to any stationary black holes.

\item Brien Nolan, Dublin City U, brien.nolan@dcu.ie

Stability of naked singularities in self-similar collapse (gr-qc/0010032)

We show that spherically symmetric self-similar space-times possessing
naked singularities are stable in the class of spherically symmetric
self-similar space-times obeying the strong and dominant energy
conditions. The discussion is restricted to space-times obeying a 'no
pure outgoing radiation' condition.

\item Richard I Harrison, U Oxford, harrison@maths.ox.ac.uk

A numerical study of the Schroedinger-Newton equation

I wish to report on a numerical study of the Schroedinger-Newton
equations, that is the set of nonlinear partial differential
equations, consisting of the Schroedinger equation coupled with the
Poisson equation.  The nonlinearity arises from using as potential
term in the Schroedinger equation the solution of the Poisson equation
with source proportional to the probability density. Penrose [1] has
suggested that the stationary solutions of the Schroedinger Newton
equation might be the 'preferred basis' of endpoints for the
spontaneous reduction of the quantum-mechanical wave-function.

I have computed stationary solutions in the spherically symmetric and the
axially symmetric cases, and then tested the linear stability of these
solutions. All solutions are unstable except for the ground state.  In the
spherically symmetric case, I have considered the general time evolution
which confirms the picture from linear theory and shows that the general
evolution leaves a lump of probability in the ground-state, while the rest
disperses to infinity. In the z-independent time evolution, initial
indications are that lumps of probability orbit around each other before
dispersing.

[1] R Penrose Phil.Trans.R.Soc.(Lond.)A 356 (1998) 1927

\item Paul Tod, U Oxford, tod@maths.ox.ac.uk

Causality and Legendrian-linking

A point $p$ in Minkowski space $M$ can be determined by its 'sky',
which is to say the set $S_p$ of null-geodesics through it in the
space $N$ of all null-geodesics. It was suggested by Penrose, and
proved in his thesis by Robert Low [1], that causal relations in $M$
are reflected by linking in $N$. Thus two points $p$ and $q$ are
time-like separated in $M$ if their skies are linked in $N$, and
space-like separated if their skies are unlinked (if they are null
separated then evidently their skies meet). Penrose also suggested
that this relationship should continue to hold for curved but, say,
globally-hyperbolic space-times $\cal{M}$. This is much harder. It was
explored by Low and later by my student Jos\'e Natario. It seems to be
true in $2+1$-dimensions but is not true in $3+1$, where one has
explicit counter-examples. Rather than give up, one can change the
question: spaces of null geodesics are contact manifolds and skies are
Legendrian submanifolds, so one can ask instead are points
causally-related iff their skies are Legendrian-linked - that is, can
they be unlinked while remaining Legendrian? There are partial answers
and various interesting developments here, and I will describe
progress on this programme.

[1] RJ Low {\em Twistor linking and causal relations} 
Class.Quant.Grav.{\bf 7}(1990) 177-187

\item  Tim Sumner, Imperial College, t.sumner@ic.ac.uk

Fundamental physics experiments in space

There is a wide interest in the UK in carrying out, so-called,
'Fundamental Physics' experiments in space.  Earlier this year the
space community produced a summary document for the PPARC SSAC.  This
talk will summarise that document, which contains suggestions for a
number of experiments to do with gravity as this is one area in which
the use of space is particularly beneficial.

\item Mike Plissi, U Glasgow, m.plissi@physics.gla.ac.uk

The GEO 600 gravitational wave detector

A number of interferometer-based gravitational wave detectors are
currently being constructed in several countries. The GEO 600 detector,
which is being built near Hannover, Germany, is a 600 m baseline
instrument that utilises a Michelson interferometric scheme. The
instrument will target the frequency band above about 50 Hz. A basic
description of the detector will be given with a report of its current
status.

\item Oliver Jennrich, U Glasgow, o.jennrich@physics.gla.ac.uk

LISA: A ESA Cornerstone mission to detect low
frequency gravitational waves.

LISA, a space-borne interferometric gravitational wave detector has
been recently approved as a ESA Cornerstone Mission. LISA makes use of
interferometry very similar to the ground-based detectors (LIGO,
VIRGO, GEO600, TAMA) but is designed to detect gravitational waves in
a much lower frequency band of 0.1 mHz to 100 mHz.

The main objective of the LISA mission is to learn about the
formation, growth, space density and surroundings of massive black
holes for which there is a compelling evidence to be present in the
centers of most galaxies, including our own.

Observations of signals from these sources would test General
Relativity and particularly black-hole theory to unprecedented
accuracy.

\item Mike Cruise, U Birmingham, amc@star.sr.bham.ac.uk

Very high frequency gravitational wave detectors

A number of theoretical models of the early Universe predict spectra of
stochastic gravitational waves rising with frequency. Such spectra
satisfy all the known observational upper limits. Detectors are needed
in the Megahertz and Gigahertz ranges to detect this radiation. A
prototype of one such detector is now in operation and the prospects
of it achieving useful sensitivities will be discussed.

\item  Edward Porter, U Cardiff, Edward.Porter@astro.cf.ac.uk

An improved model of the gravitational wave flux for inspiralling black 
holes

While the orbital energy for an inspiralling binary is known exactly for 
both the Schwarzchild and Kerr cases, an exact expression for the 
gravitational wave flux remains elusive.  All current analytical models rely 
on a Post-Newtonian expansion.  This has given us a Taylor expansion for 
the flux in the test-mass case to $v^11$ for the Schwarzchild case, and to 
$v^8$ for the Kerr case.  The problem with the Taylor approximation is the
slow rate of convergence at various approximations.  It has been shown 
that using Pade Approximation gives a better convergence for the flux in the 
Schwarzchild case.  In this work I propose a method for improving the 
convergence of the flux in the Schwarzchild case by using a modified Pade 
Approximation and extend the previous work to the Kerr case.  The most 
interesting result from the Kerr case is that we may be able to closely 
model a Kerr system with some real value of the spin parameter 'a' with a 
Pade Approximation using a 'wrong' value of 'a'.  We also provide scaling 
laws at various approximations to recover the true spin of the system from 
the Pade Approximation.

\item Anna Watts, U Southampton

Neutron stars as a source of gravitational waves

We examine the evolution of the r-mode instability for a magnetized 
neutron star
accreting large amounts of remnant matter in the immediate aftermath of
the supernova.  We discuss the implications for neutron star spin rate and
gravitational wave signal.

\item John Miller, SISSA / U Oxford, miller@sissa.it

Non-stationary accretion onto black holesx

Update on our project for using computer simulations to investigate
different pictures for non-stationary accretion onto black holes. This has
relevance for explaining observed time-varying behaviour of galactic X-ray
sources and AGN and, possibly, the formation of jets.

\item Uli Sperhake, U Southampton, us@maths.soton.ac.uk

A new numerical approach to non-linear oscillations of neutron stars

Radial oscillations of neutron stars are studied by decomposing the
fundamental variables into a background contribution
(taken to be the static TOV-solution) and time dependent perturbations.
The perturbations are not truncated at some finite order,
but are evolved according to the fully nonlinear evolution equations,
which can be written in quasi linear form in our case. The seperation
of the background allows us to study oscillations over a wide range of
amplitudes with high accuracy. We monitor the onset of nonlinear
effects as the amplitude is gradually increased. Problems
encountered at the surface in any Eulerian formulation, i.e. the
singular behavior of the equations in the nonlinear as well as the
linearised case and its impact on our results is briefly discussed.

\item Philippos Papadopoulos, U Portsmouth, philippos.papadopoulos@port.ac.uk

Non-linear black hole oscillations (gr-qc/0104024)

The dynamics of isolated black hole spacetimes is explored in
the non-linear regime using numerical simulations. The geometric setup is
based on ingoing light cone foliations centered on the black hole. The
main features of the framework and the current status of the computations
will be presented.

\item Felipe Mena, Queen Mary and Westfield College, F.Mena@qmw.ac.uk

Cosmic no hair: second order perturbations of de Sitter universe

We study the asymptotic behaviour of second order perturbations in a
flat Friedmann-Roberstson-Walker universe with dust plus a
cosmological constant, a model which is asymptotically de Sitter.  We
find that as in the case of linear perturbations, the nonlinear
perturbations also tend to constants, asymptotically in time. This
shows that the earlier results concerning the asymptotic behaviour of
linear perturbations is stable to nonlinear (second order)
perturbations.  It also demonstrates the validity of the cosmic
no-hair conjecture in such nonlinear inhomogenuous settings.

\item Kostas Glampedakis, U Cardiff, Costas.Glampedakis@astro.cf.ac.uk

Scattering of scalar waves by rotating black holes (gr-qc/0102100)

We study the scattering of massless scalar waves by a Kerr black hole, by 
letting plane monochromatic waves impinge on the black hole. We calculate
the relevant scattering phase-shifts using the Pr\"{u}fer phase-function
method, which is computationally efficient and reliable also for high 
frequencies and/or large values for the angular multipole indices (l,m).
We use the obtained phase-shifts and the partial-wave approach to determine
differential cross sections and deflection functions. Results for off-axis
scattering (waves incident along directions misaligned with the black hole's
rotation axis) are obtained for the first time. Inspection of the off-axis
deflection functions reveals the same scattering phenomena as in Schwarzschild
scattering. In particular, the cross sections are dominated by the glory
effect and the forward (Coulomb) divergence due to to the long-range nature
of the gravitational field. In the rotating case the overall diffraction
pattern is ``frame-dragged'' and as a result the glory maximum is not observed
in the exact backward direction.  We discuss the physical reason for this
behaviour, and explain it in terms of the distinction between prograde and
retrograde motion in the Kerr gravitational field. Finally, we also discuss
the possible influence of the so-called superradiance effect on the scattered
waves. 

\item Reinhard Prix,  U Southampton, rp@maths.soton.ac.uk

Covariant multi-constituent hydrodynamics (gr-qc/0004076)

I will discuss the covariant formulation of hydrodynamics derived from
a "convective" variational principle by Carter. This approach allows a
convenient generalisation to several interacting fluids (incorporating
the effect known as "entrainment") and to superfluids. Such a framework
is therefore very well suited to neutron star applications, some of
which I will briefly describe here. 

\item Ian Jones, U Southampton, dij@maths.soton.ac.uk

Gravitational waves from freely precessing neutron stars (gr-qc/0008021)

The free precession of neutron stars has long been cited as a possible
source of detectable gravitational radiation.  In this talk we will
examine the problem of calculating the gravitational radiation reaction on
a star, which we model as an elastic shell containing a fluid core.  We
will conclude by assessing the likely gravitational wave amplitudes of
precessing neutron stars in our Galaxy.

{\bf The following talks were also scheduled, but the speakers had to
cancel:}

\item  Bernard S. Kay, U York, bsk2@york.ac.uk

New paradigm for decoherence and for thermodynamics, new
understanding of quantum black holes (hep-th/9802172, hep-th/9810077) 

I outline my recent proposed new explanation for decoherence and
for entropy increase based on my new postulate that the "quantum
gravitational field is unobservable" and on my related new postulate
that "physical entropy is matter-gravity entanglement entropy".  I
also recall how this proposal offers a resolution to a number of
black-hole puzzles, including the "information loss puzzle".

\item Alberto Vecchio, U Birmingham, vecchio@aei-potsdam.mpg.de

Searching for binary systems undergoing precession with GEO and LIGO
(gr-qc/0011085)

The search for binary systems containing rapidly spinning black holes
poses a tremendous computational challange for the data analysis of
gravitational wave experiments.  We present a short review of our
present understanding of the key issues and discuss possible
strategies to tackle efficiently the problem.

\end{enumerate}
\end{document}